\documentclass[a4paper]{jpconf}
\usepackage{graphicx}
\newcommand\eq[1]{\begin{eqnarray}#1\end{eqnarray}}

\newcommand{\mb}{\mathbf}

\begin{document}
\title{Kraichnan like model of turbulence with stratification}
\author{Heikki Arponen}
\address{University of Helsinki, Dept. of Mathematics and Statistics,P.O. Box 68 (Gustaf H\"allstr\"omin katu 2b)
FI-00014 University of Helsinki}
\ead{heikki.arponen@helsinki.fi}
\begin{abstract}
A generalization to the Kraichnan gaussian, white noise in time model of turbulent velocity field is proposed. The generalization is designed to take into account the effects of stratification in the atmospheric boundary layer by introducing different scaling exponents for horizontal and vertical velocity components, leading to strong vertical anisotropy. The inevitable compressibility of the model leads one to take into account also the density statistics of the fluid, which will immediately lead to a boundary layer like structure in the flow statistics and non zero time correlations. Possible applications for passive scalar, tracer and inertial particle statistics and their anomalous scaling behavior will be discussed.
\end{abstract}
\section{Introduction}
Compared to homogeneous, isotropic turbulence, perhaps the most significant difference in a stably stratified atmospheric boundary layer is the strong vertical anisotropy manifesting as a $\propto k^{-3}$ mesoscale spectrum (corresponding to $\propto r^2$ spatial behavior in the structure function)  of the vertical velocity components, as opposed to the usual Kolmogorov-Obukhov $\propto k^{-5/3}$ spectrum of the horizontal components (see e.g. \cite{brethouwer} and references therein). The vertical velocities are "smoothened" by the effect of gravity pulling the atmosphere toward the ground. Recently, much understanding on the nature of intermittency and anomalous scaling (or multiscaling) has been acquired by the so called Kraichnan model, where one assumes the turbulent velocity statistics to be known and completely determined by a gaussian, white in time random velocity field (see e.g. \cite{falkovich} for a review). The purpose of the present work is to generalize the Kraichnan model to mimic the behavior of stratified flows. The ultimate goal of the project would be to determine how the stratification affects the anomalous scaling exponents of the passive scalar, and to study possible clustering of inertial particles.
\section{Kraichnan model}
The Kraichnan model is a gaussian, mean zero, white in time random field which can be completely defined via the pair correlation function
\eq{\langle v_\mu (t,\mb x) v_\nu (t',\mb x')\rangle \doteq \delta(\Delta t) D_{\mu \nu}(\Delta \mb x),}
where $\delta (\Delta t) \doteq \delta(t-t')$ is the Dirac delta function and $D_{\mu \nu}$ is an incompressible tensor field, which can be defined e.g. as
\eq{D_{\mu \nu} (\mb r) \doteq D_0 \int \! \! d^3 \mb p \frac{\delta_{\mu \nu} -p_\mu p_\nu /p^2}{\left(p^2 +L^{-2} \right)^{3/2+\xi/2}} e^{i \mb p \cdot \mb r}. \label{Dmunu}}
The constant $L$ is the integral scale of the flow and $0 < \xi <2$ describes the spatial roughness of the velocity field and appears as a scling exponent in the velocity structure function in the limit of infinite $L$ as
\eq{\langle \left| \mb v (t, \mb x)-\mb v (t', \mb x')  \right|^2 \rangle \propto \delta (\Delta t) |\Delta \mb x|^\xi.}
One should note that although the infinite $L$ limit of the structure function is finite, the tensor field of (\ref{Dmunu}) diverges as $L^\xi$.
\\ \\
The passive scalar evolves according to the equation
\eq{\partial_t \theta -\kappa \Delta \theta + \bar v \cdot \nabla \theta = f,}
where $f \doteq f\left( t, \mb r \right)$ is usually defined as a gaussian, white in time forcing, defined as
\eq{\langle f\left( t, \mb r \right)f\left( t', \mb r' \right) \rangle \doteq \delta (\Delta t) \mathcal C \left( \Delta \mb r/L_f \right)}
with a forcing scale $L_f$. In the case of an isotropic, large scale forcing, i.e. $\mathcal C \to \mathcal C (0)$ as $L_f \to \infty$, the passive scalar equal time structure functions scale as
\eq{\langle \left[\theta(t,\mb x+\mb r)-\theta(t,\mb x) \right]^N \rangle \propto r^{\frac{N}{2}(2-\xi)-\Delta_N}}
with $\Delta_N = \frac{N(N-2)}{2(d+2)}\xi+\mathcal O (\xi^2)$ for small $\xi$. The simple gaussian model of turbulence has therefore lead to an intermittency effect manifesting as multiscaling of the passive scalar structure functions.
\section{Modified Kraichnan model with stratification}
The idea of the present work is to study the effect of stratification in the context of a Kraichnan like turbulence model. For example, one would like to determine if and how the stratification affects the anomalous scaling exponents $\Delta_N$ above. We define a stratified generalization of the Kraichnan model as
\eq{D_{i \nu} \doteq D_0 \int \! \! d^3 p \frac{P_{i\nu}(\hat p)  e^{-l p}}{\left(p^2 +L^{-2} \right)^{3/2+\xi/2}}e^{i \bar p \cdot \bar r} =D_{\nu i} }
for the non-vertical components with $i \in {1,2}$, $\mu \in {1,2,3}$ and $r_3$ is the vertical component. The vertical component is taken as
\eq{D_{33} \doteq D_0 L^{\xi-\eta} \int \! \! d^3 p \frac{P_{33}(\hat p)  e^{-l p}}{\left(p^2 +(\alpha L)^{-2} \right)^{3/2+\eta/2}}e^{i \bar p \cdot \bar r}.}
Here $0 \leq \xi < \eta \leq 2$ are the horizontal and vertical scaling exponents, $\alpha \leq 1$ is the aspect ratio of vertical to horizontal integral scales and the term $L^{\xi-\eta}$ term in the prefactor ensures correct dimensionality. In an atmospheric boundary layer one would have $\eta \approx 2$, but we will keep $\eta$ as a free parameter. Also included is an ultraviolet cutoff term $\propto e^{-l p}$, where the length scale $l$ corresponds to a viscous dissipative scale. It is often omitted in the Kraichnan model, since one is usually interested in scales larger than the viscous dissipation and since the model is finite as $l$ is taken to zero. Here the viscous scale $l$ must be kept finite (for now) because the model is not incompressible due to the different horizontal and vertical scaling. Specifically, the quantity $\langle (\nabla \cdot \mb v)^2\rangle$ diverges as $l$ approaches zero. The diffusivity ratio behaves as $D_{33}/D_{ii} \propto \alpha^\eta$, which implies vertically inhibited diffusion for small aspect ratios, as is known to occur in real stratified turbulence (see e.g. \cite{vincent}).\\ \\
%

Finite compressibility in the model forces one to reconsider the role of density in the model, i.e. it is no longer possible to simply assume a constant density. Instead we take the density as another random field beside the velocity field. This is most conveniently achieved by rewriting the mass continuity equation
\eq{\partial_t \rho + \nabla \cdot \left(\bf{v} \rho \right )=0\label{masscont}}
in terms of the logarithm of the density (henceforth called the log-density) $\chi \doteq \log \left( \rho/\rho_0 \right)$, where $\rho_0$ is some reference density. The reason can be seen from the passive scalar equation, which in the case of nonconstant density reads
\eq{\partial_t \left( \rho \theta \right ) +
\nabla \cdot \left(\bf{\bar v} \rho \theta \right )- \kappa \nabla \cdot \left(\rho \nabla \theta \right )=0.}
As usual, the above equations should be interpreted as a stochastic PDE of Stratonovich (mid point) type \cite{zinn-justin}. Taking this carefully into account and going from Stratonovich to It\^o prescription (see e.g. the appendix A of \cite{arponen}), using the continuity equation (\ref{masscont}) and finally taking the limit of vanishing molecular diffusivity $\kappa$, we obtain
\eq{\partial_t \theta + \left[ v_\mu + D_{\mu\nu}(0)\partial_\nu \chi \right] \partial_\mu \theta - \frac{1}{2}\widehat D_0 \theta = 0,}
with the notation $\widehat D_0 \doteq D_{\mu\nu}(0) \partial_\mu \partial_\nu$. The passive scalar is therefore advected by an effective velocity field composed of the Kraichnan velocity field $\mb v$ and a gradient of the log-density $\chi$. The equation for the log-density follows by similar use of the rules of stochastic calculus from eq. (\ref{masscont}), giving
\eq{\partial_t \chi+\nabla \cdot \mb v + \mb v \cdot \nabla \chi -\frac{1}{2}\widehat D_0 \chi + \gamma =0,}
with notation $\gamma \doteq \langle (\nabla \cdot \mb v)^2\rangle = -D_{\mu\nu,\mu\nu}(0)$ (symbols after a comma denote derivatives).\\ \\

\begin{figure*}

\begin{center}

\includegraphics[scale=.8]{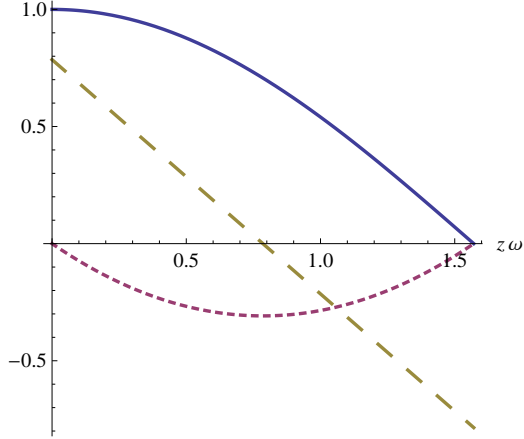}
\caption{\label{densities} Plots of mean density (solid), mean log-density (small dashes) and vertical mean effective velocity (large dashes).}
\end{center}
\end{figure*}

Using the above SPDE, we write the steady state equations for the mean log-density $g(z) \doteq \langle \chi \rangle$ (which we assume to depend only on the vertical coordinate $z=x_3$),
\eq{\widehat D_0 g = 2 \gamma,}
which can be readily solved with suitable boundary conditions, yielding
\eq{\langle \chi \rangle = g(z) = \frac{1}{2}\omega z \left( \omega z-\frac{\pi}{2} \right),\label{log-density}}
where $\omega^2 \doteq 2 \langle (\nabla \cdot \mb v)^2\rangle /\langle \mb v^2\rangle \propto \alpha^{-\eta} \left( l/L \right)^{\xi} l^{-2}$. The solution will therefore tend to zero as $L$ approaches infinity. It is perhaps more instructive to consider the mean density rather than the mean log-density. The solution for mean density is
\eq{\langle \rho \rangle = \rho_0 \cos \left( \omega z \right),}
which leads one to consider the "height" of the boundary layer as $h\doteq \pi /2\omega \propto L^{\xi/2}$. The gradient of the mean log-density (\ref{log-density}) implies that the mean effective velocity has a nonzero vertical component, pointing towards the mean height. This implies that passive quantities will exhibit mean values concentrated at height $h/2$. The situation is illustrated in Fig. (\ref{densities}). Clearly this is not a completely realistic density profile, as one might expect since the velocity correlation function does not exhibit any height dependence in this approximation.\\ \\

The equation for the flux like correlator $\mathcal F_\mu \left( t-t', \mb x -\mb x' \right)\doteq \langle \chi (t,\mb x) v_\mu (t', \mb x')\rangle$ is
\eq{\partial_t \mathcal F_\mu = - \delta (t-t') \left[  D_{\mu\nu,\nu} + g'(z) D_{\mu 3} \right] + \frac{1}{2} \widehat D_0 \mathcal F_\mu.}
Since it is the gradient of the log-density that appears e.g. in the passive scalar equation, we take a derivative of the solution, which at leading order as $L \to \infty$ is approximately
\eq{\langle \partial_\mu \chi(t, \mb x) v_i(t',\mb x')\rangle\!\!\!\!\!\!\!\!\! &&= - \theta (\Delta t) e^{\frac{1}{2}\Delta t\widehat D_0}  \partial_\mu\left[ D_{3 i} g'(z)+D_{i\nu,\nu} \right] \\ && \approx \theta (\Delta t) \left[  c_{\mu\nu}^1 +c_{\mu\nu}^2 (r/L)^2+c_{\mu\nu}^3 \Delta t/L^{2-\xi}\right],}
where the $c^i$ are some anisotropic unit tensors and $r = |\Delta \mb x|$. If we further assume that $\Delta t /L^{2-\xi} \gg 1$, we obtain $\langle \chi v_i'\rangle \propto \theta (\Delta t) (\Delta t)^{-(5-\xi)/2}$, i.e. the long time time correlation is a power law. Similar result for the vertical velocity component can be derived but with $\xi$ replaced by $\eta$. \\ \\

For the equal time two point function $G_0 (\mb x,\mb x')\doteq \langle \chi (t,\mb x) \chi (t, \mb x') \rangle$ we have
\eq{\frac{1}{2} D_{\mu\nu}(0) \left( \partial_\mu \partial_\nu + 2 \partial_\mu \partial_\nu' + \partial_\mu' \partial_\nu'\right)G_0 - d_{\mu\nu} \partial_\mu \partial_\nu' G_0 -D_{\mu\nu,\mu\nu} -\gamma \left( g(z)+g(z') \right)= 0.}
At leading order the last term vanishes. We can then assume homogeneity, which results also in the vanishing of the first term, yielding the leading order solution (again with gradients of $\chi$)
\eq{\langle \partial_\mu \chi (t,\mb x) \partial_\nu\chi (t, \mb x') \rangle &&\approx - \partial_\mu\partial_\nu\widehat d^{-1}D_{\alpha\beta,\alpha\beta} (\mb x - \mb x') \\ &&\approx c_{\mu \nu}^4 r^{-2} ,}
with another anisotropic unit tensor and notation $\widehat d = d_{\mu\nu} \partial_\mu \partial_\nu = \left( D_{\mu\nu}(0)-D_{\mu\nu} \right)\partial_\mu \partial_\nu$. We've also assumed $r \gg l$, hence the divergence at $r=0$. \\ \\

The equation for steady state unequal time two point function $G(t-t',\mb x, \mb x') \doteq \langle \chi (t,\mb x) \chi (t', \mb x') \rangle$ is in turn,
\eq{\partial_t G = \frac{1}{2}\widehat D_0 G - \gamma g(z'),\ \ \ \ \ t>t'.}
The solution, again at leading order, is
\eq{\langle \partial_\mu \chi (t,\mb x) \partial_\nu \chi (t', \mb x') \rangle &&\approx - e^{\frac{1}{2} |\Delta t| \widehat D_0} \partial_\mu \partial_\nu G_0 \\ &&\approx c_{\mu\nu}^5 L^{-\xi} \frac{1}{|\Delta t|}}
with another anisotropic unit tensor.
\section{Concluding remarks}
The next step in the process would naturally be to consider e.g. the passive scalar and tracer or inertial particles in a flow defined by the above model. From the approximative expressions above, one can already gain some insight into some properties of these passive quantities. First, since there is now a mean effective velocity field pointing toward the height $h/2$, this is where one may expect the particles and passive scalar fields to be concentrated. Second, it was demonstrated that the log-density gradient field exhibits a time-like correlation, as opposed to the white noise in time correlation of the Kraichnan model. As is well known, the phenomenon of clustering of inertial particles requires a non zero correlation time on order for the particles to have time to traverse to the edges of eddies. It is therefore reasonable to expect that such clustering may occur for this model, simply due to the anisotropic density fluctuations. Also, a common critique toward extensions of the Kraichnan model to models with time like correlations is that they violate Galilean invariance. Here such a problem does not arise since the log-density field is simply determined by the mass continuity equation, which is of course Galilean invariant by construction.
\ack
The author would like to thank Antti Kupiainen for useful discussions concerning the project. This work was supported by the Magnus Ehrnrooth foundation.
\section*{References}

\end{document}